# THERMODYNAMIC LIMIT OF THE

# GINZBURG LANDAU EQUATIONS

P.COLLET

Centre de Physique Théorique
Laboratoire CNRS UPR 14
Ecole Polytechnique
F-91128 Palaiseau Cedex (France)

**Abstract**: We investigate the existence of a global semiflow for the complex Ginzburg-Landau equation on the space of bounded functions in infinite domain. This semiflow is proven to exist in dimension 1 and 2 for any parameter values of the standard cubic Ginzburg-Landau equation. In dimension 3 we need some restrictions on the parameters but cover nevertheless some part of the Benjamin-Feijer unstable domain.

# I INTRODUCTION.

Non linear parabolic equations appear quite often in the description of the time evolution of important physical systems. In systems with finite spacial extension, these equations must be supplemented by boundary conditions in order to give rise to well defined (non linear) semi-flows of evolution. When the domain is large with respect to the size of the typical structures, physicists in particular have used since a long time an infinite domain approximation. This is quite convenient for explicit computations since one can use for example Fourier transforms. From a rigorous point of view this raises the question of existence regularity and boundedness of solutions for non linear parabolic PDE in unbounded space domains. The goal of this paper is to prove some results in this direction for a family of PDE which appear quite often as normal forms in the study of the bifurcations with continuous spectrum (see [N.W.], [S.], [C.E.1], [vH.], [Mi.], [Sc.], [I.M.D.], [I.M.], [H.M.]). These are the so called complex Ginzburg Landau equations ($CGL_q$ for short) which are given by

$$\partial_t u = (1+i\alpha)\Delta u + P(u, \overline{u}) - (1+i\beta)|u|^{2q}u \qquad (CGL_q)$$

where $u$ is a complex valued function of a space variable $x \in \mathbb{R}^d$ and of a time variable $t \in \mathbb{R}^+$. $q$ is a positive integer, $\alpha$ and $\beta$ are real numbers, and $P$ is a complex polynomial in two variables of degree at most $2q$. It will be convenient to denote by $NL(u)$ the quantity

$$NL(u) = P(u, \overline{u}) - (1+i\beta)|u|^{2q}u .$$

For simplicity of notations we will only treat below the case where $P(u, \overline{u}) = u$ although the general case is similar.

The main problem discussed in this paper is whether these equations define a semi-flow in $L^\infty(\mathbb{R}^d)$ which is global in time. Note first that the Cauchy problem in this space is not difficult to handle and we have existence and regularity of the solution for a small time interval which depends on the $L^\infty$ norm of the initial data (see [Pa.] [H.] and Theorem 7 below). The global time problem requires some a priori estimates on the solutions. For bounded domains or for solutions in Sobolev spaces there is a vast amount of results in the literature (see [Y.] [B. et al.] and [D.H.1] for recent results).

A major difference of the present work with previously published results is that we will be able to treat functions with no specified behavior at infinity (except of course that they are bounded). This will include for example the complex exponentials $e^{ikx}$ often considered in the Physical literature. This is particularly important since many interesting particular solutions have been constructed like stationary solutions, fronts, slugs etc. (see [C.E.2], [D.H.2], [I.M.] for some examples). One would like of course to put all these solutions into a unified framework to be able in particular to discuss the fundamental problem of stability of these solutions. We recall that linear stability for example requires to investigate the spectrum of a linearized operator and this spectrum may depend crucially on the function space (see [C.E.2], [I.M.]).

The methods presented below can be extended in many directions: to other type of equations (for example the modified Ginzburg Landau equation [D.H.T.]), vector valued functions $u$, variable coefficients etc. We will refrain even to attempt listing the possible extensions and concentrate on showing how the method works in the case of $CGL_q$.

A related question has to do with the approximation of large (but finite) systems by infinite ones. The idea has one of it's origin in the thermodynamic limit of Statistical Mechanics where one argues that large systems behave mostly like infinite ones. In our present context, a large system is a system in a domain whose size is much larger than the typical size of the local structures of the solutions. There are many such physical situations, for example the clouds (and even some big storms) are much smaller than the size of earth, the waves in the sea have a wavelength much smaller than the size of the ocean etc. As we mentioned before in a bounded domain the existence of the semi-flows solving the above equations is well understood. However the methods developed below produce bounds which are uniform in the (large) size of the system if one is far away form the boundary. These bounds are also independent of the boundary conditions in the thermodynamic limit. This requires of course a reasonable notion of thermodynamic limit, that is to say a class of domains with a geometry of the boundary not too involved. We recall that extensive bounds for energy estimates were already obtained in [G.H.].

In the case of bounded domain, the main tool to establish a priory estimates is to use energy (or Lyapunov) functionals. In the present case this method cannot be used directly because these functionals diverge and it is not clear how one can renormalize them for the general class of functions we are considering. Some infinite domain problems with given "boundary condition at infinity" can be treated by this method ([D.H.1], [Y.]) eventually after renormalization of the Lyapunov functional (substraction of an adequate function).

The basic idea to treat the infinite domain problem will be to use local energy estimates and this will be done by introducing an adequately chosen smooth space cut-off in the functional. There is some liberty in the choice of this cut-off and we will use several of them. A simple example is the function

$$\varphi(x) = \frac{1}{(1+a^2|x-x_0|^2)^{1+d/2}} ,$$

where $a$ is a positive real number, and $x_0$ is a fixed point in $\mathbb{R}^d$. The number $a$ will in general be chosen small, and varying $x_0$ over all $\mathbb{R}^d$ will provide uniform estimates. This choice of $\varphi$ is particularly useful because it is an integrable function, and we have the following estimate

$$|\nabla \varphi(x)| \leq \mathcal{O}(1) a \varphi(x) , \tag{1.1}$$

which allows to control the new terms in the energy estimates. This method was previously used in [C.E.1] to study the real Ginzburg Landau and the Swift Hohenberg equation in dimension one. It is however of limited use for systems in large but finite domain. Similar cut off functions were also used in [B.V.].

In the present paper we will extend these results to higher dimension using different techniques. Note in particular that the function $\varphi$ is not of compact support, and therefore the energy estimates will be only partially localized in space. In order to make contact with the large but finite volume problems where boundary conditions are still present we will use cut-off functions with compact support. However these functions will depend of the particular problem. For $\text{CGL}_q$ we will use any nonnegative $C^2$ function with compact support $\varphi$ and which instead of the estimate (1.1) satisfies the estimate

$$\frac{|\nabla\varphi|^2}{\varphi} \leq \mathcal{O}(1)\varphi^{1/(q+1)} \tag{1.2}$$

A simple example is given by

$$\varphi_q(x) = \begin{cases} (1 - |x - x_0|^2)^{2+2/q} & \text{if } |x - x_0| < 1, \\ 0 & \text{otherwise.} \end{cases}$$

In the case of $CGL_1$ in dimension 2 we will have to us a more adapted cut-off function that will be defined later on.

In all cases it is also important to obtain bounds which are translation invariant. We will use in fact a one parameter family of cut-off functions which are translates of the initial one by a fixed vector $x_0$ (namely the function $y \to \varphi(y - x_0)$). The goal is to obtain estimates which are independent of $x_0$, and this uniformity will be the source of the $L^\infty$ estimate. In order to simplify the notation, we will not mention explicitly the dependence on the translation parameter $x_0$.

The paper is organized as follows. In the next section we establish the local energy estimates, and related estimates concerning the (local) $L^2$ norm of the gradients. This will work quite simply in some ranges of parameters $\alpha$ and $\beta$. For other parameters we will have to use Gagliardo-Nirenberg inequalities and this will restrict the values of the dimensions. In the third section we use these estimates to prove the semi-flow property using a local integral form of the equations. We will also discuss some consequences for bounded domains.

Before proceeding we introduce some general notations. First all the integrals will be with respect to the Lebesgue measure (in the appropriate dimension) and we will not write explicitly this measure in the integrals. We will also not write explicitly the arguments of the integrand when it is obvious. $|\ |$ will denote the norm of a vector (absolute value for a number), and $\bullet$ will denote the scalar product. $\Re$ will denote the real part, and $C$ will denote various positive constants which are independent of $u$ and may take different values in different statements (one may of course consider once for all a large enough constant which can be used in all the statements simultaneously). Finally we will denote by $\|\ \|_{a,b,O}$ the Sobolev norm in the set $O$ which is the sum of the $L^b$ norms of the partial derivatives up to order $a$, while $\|\ \|_{b,O}$ will denote the usual $L^b$ norm.

## II THE LOCAL ENERGY ESTIMATES.

We start by proving the basic energy inequality which can be formulated as follows.

**Theorem 1.** *For any $CGL_q$ in any dimension, and for any nonnegative function $\varphi$ $C^2$ with compact support and satisfying (1.2) there is a positive constant $C$ such that for any solution of the Cauchy problem in $L^\infty$*

$$\partial_t \int \varphi |u|^2 \leq C - \int \varphi |u|^2 - \int \varphi |\nabla u|^2 - \int \varphi |u|^{2q+2} .$$

**Proof.** Using integration by parts we have easily

$$\frac{1}{2}\partial_t \int \varphi |u|^2 =$$

$$- \int \varphi |\nabla u|^2 - \Re \left( \int \overline{u} \nabla \varphi \bullet \nabla u \right) - \Re \left( \int (1+i\alpha)\varphi \overline{u} P(u, \overline{u}) \right) - \int \varphi |u|^{2q+2} .$$

We now add and substract to the right hand side the quantity $\int \varphi |u|^2$, and we obtain

$$\frac{1}{2}\partial_t \int \varphi |u|^2 \leq - \int \varphi |u|^2 - \int \varphi |\nabla u|^2 +$$

$$\mathcal{O}(1) \int |\nabla \varphi||u||\nabla u| + \mathcal{O}(1)(1+|\alpha|) \int \varphi(1+|u|^{2q})|u| - \int \varphi |u|^{2q+2} + \int \varphi |u|^2 .$$

By polarization we have easily using (1.2)

$$-\frac{1}{2} \int \varphi |\nabla u|^2 + \mathcal{O}(1) \int |\nabla \varphi||u||\nabla u| - \frac{1}{4} \int \varphi |u|^{2q+2} + \int \varphi |u|^2$$

$$\leq \mathcal{O}(1) \int \frac{|\nabla \varphi|^2}{\varphi} |u| - \frac{1}{2} \int \varphi |u|^{2q+2} + \int \varphi |u|^2 \leq \mathcal{O}(1) .$$

The Theorem now follows easily by polarization of the remaining terms.

All the following estimates will depend on a positive function $y(t)$ given by

$$y(t) = \max(2C, \sup_{x_0} \int \varphi(x-x_0)|u(x,t)|^2 \, dx) .$$

We observe that (assuming $\varphi$ is not identically zero), that for any compact set $K$, there is a constant $C_K$ such that under the hypothesis of Theorem 1, we have

$$\int_K |u(x,t)|^2 \leq C_K y(t) .$$

We now deduce from Theorem 1 the main property of the function $y$.

**Corollary 2.** *Under the hypothesis of Theorem 1, the function $y$ is non increasing. It decreases if it is larger than $2C$, and reaches this value after a finite time (provided of course we know that the solution remains bounded at least up to this time). Moreover for any time interval $[T - \tau, T]$ ($\tau < T$) where we have existence of the solution and it's gradient in $L^\infty$ we have the bound*

$$\int_{T-\tau}^{T} dt \left( \int \varphi |\nabla u|^2 + \int \varphi |u|^{2q+2} \right) \leq C\tau + y(T - \tau) .$$

**Proof.** This is an immediate consequence of Theorem 1.

Note that the above bound is independent of $x_0$. This implies in particular that we obtain bounds for the integrals on any fixed compact set which are uniform in time. All subsequent bounds will be given in terms of the non increasing function $y$. This is why we will obtain at the end a bound asymptotic in time which is independent of the initial condition.

We now come to the estimates concerning the gradient of the function $u$. We first start with the simple case.

**Theorem 3.** *For any function $\varphi$ nonnegative, $C^2$ with compact support, satisfying (1.2) and such that $|\nabla \varphi|^2 / \varphi$ satisfies also these hypotheses, and for any $CGL_q$ in any dimension such that*

$$\alpha \beta > 0 , \text{ or } \sqrt{1 + \alpha^2} < 1 + 1/q , \text{ or } \sqrt{1 + \beta^2} < 1 + 1/q ,$$

*there are positive constants $C$, $\rho$ and $\theta$ such that for any regular solution of the Cauchy problem in $L^\infty$ (also for the derivatives up to order four), then the function*

$$z(t) = \int \varphi |\nabla u|^2 + \rho \int \varphi |u|^{2q+2} + \theta \int \left( \varphi + \frac{|\nabla \varphi|^2}{\varphi} \right) |u|^2$$

*satisfies*

$$\partial_t z \leq C .$$

**Proof.** Using integration by parts and polarization as in the proof of Theorem 1, we obtain easily for any positive number $\epsilon$

$$\partial_t \int \varphi |u|^{2q+2} \leq -2(q+1)^2 \int \varphi |u|^{2q} |\nabla u|^2 - 2(q+1) \int \varphi (|u|^{4q+2} - |u|^{2q+2})$$

$$-2q(q+1) \int \varphi |u|^{2q-2} \Re((1 + i\alpha)\overline{u}^2 (\nabla u)^2) + \mathcal{O}(1)\epsilon^{-1} \int \frac{|\nabla \varphi|^2}{\varphi} |\nabla u|^2 + \epsilon \int \varphi |u|^{4q+2}.$$

Similarly we have for any positive number $\epsilon$

$$\partial_t \int \varphi |\nabla u|^2 \leq -2 \int \varphi |\Delta u|^2 - 2(q+1) \int \varphi |u|^{2q} |\nabla u|^2 + 2 \int \varphi |\nabla u|^2$$

$$-2q \int \varphi |u|^{2q-2} \Re((1+i\beta)u^2(\nabla \overline{u})^2) + \epsilon \int \varphi |\Delta u|^2 + \mathcal{O}(1)\epsilon^{-1} \int \frac{|\nabla \varphi|^2}{\varphi} |\nabla u|^2 \ .$$

We now observe that in the two inequalities appear similar terms, namely

$$A = -2q(q+1) \int \varphi |u|^{2q-2} \Re((1+i\alpha)\overline{u}^2(\nabla u)^2) \ .$$

and

$$B = -2q \int \varphi |u|^{2q-2} \Re((1+i\beta)u^2(\nabla \overline{u})^2) \ ,$$

If $\alpha\beta > 0$, we have

$$\frac{\beta A + \alpha(q+1)B}{\beta + \alpha(q+1)} \leq 2q(q+1) \frac{\beta + \alpha}{\beta + \alpha(q+1)} \int \varphi |u|^{2q} \nabla \overline{u}|^2) \ .$$

The result now follows easily by considering the above convex combination of the two differential inequalities. If $\sqrt{1+\alpha^2} < 1 + 1/q$, we have already

$$A < 2q^2 \int \varphi |u|^{2q} |\nabla u|^2 \ ,$$

and the estimate follows again by taking a convex combination of the two differential inequalities with a small coefficient for the second one. The case $\sqrt{1+\beta^2} < 1 + 1/q$ is treated similarly.

We now come to the case where there is no restriction on $\alpha$ and $\beta$. Note however that the previous estimate already covers some interesting regions of parameters like $\alpha\beta < -1$ which corresponds to the Benjamin Feijer instability for $q = 1$ (see [L.N.P.]). In the general case we will have to use the Sobolev inequalities and the results will depend crucially on the dimension and on the number $q$. Moreover we will use a cut-off function $C^\infty$ with compact support $\psi$ which is defined by

$$\psi(x) = \begin{cases} e^{-1/(1-d(x,x_0)^2)} & \text{if } d(x,x_0) < 1, \\ 0 & \text{else.} \end{cases}$$

As for the function $\varphi$, this is a convenient and simple choice among many possibilities. As explained before, the argument $x_0$ will not be mentioned explicitly but is crucial to obtain local estimates valid on the whole line. We will also choose once for all a function $\zeta$ which is non negative, $C^\infty$ with compact support, satisfies (1.2) and is identically equal to one on the ball of radius one centered at the origin (as above we will also use it's translates without further notice).

**Theorem 4.** *For $CGL_1$ and $CGL_2$ in dimensions 1, and for $CGL_1$ in dimension 2 there are positive constants $C$, $c$ and $\rho$ such that for any regular solution of the Cauchy problem in $L^\infty$ (also for the derivatives up to order 4), the function $z(t)$ defined by*

$$z(t) = \int \psi |\nabla u|^2 + \int \psi |u|^{2q+2} + \rho \int \zeta |u|^2$$

*satisfies*
$$\partial_t z \leq C + cz \int \zeta |u|^4 .$$

**Proof.** We start with the one dimensional case which is easier. As in the proof of Theorem 3 we obtain easily
$$\partial_t \left( \int \psi |u'|^2 + \int \psi |u|^{2q+2} \right) \leq$$
$$C - \int \psi |u''|^2 - \int \psi |u|^{4q+2} + \mathcal{O}(1) \int \frac{\psi'^2}{\psi} |u'|^2 + \mathcal{O}(1) \int \psi |u'|^2 |u|^{2q} .$$

If $B$ denote the ball (interval) of radius one centered at $x_0$, we have from the Sobolev inequalities (see [F.])
$$\|u^{2q}\|_{\infty,B} \leq \mathcal{O}(1)(\|u'\|_{2,B} \|u\|_{4q-2,B}^{2q-1} + \|u\|_{2q,B}^{2q}) \leq \mathcal{O}(1)(1 + \|u'\|_{2,B}^2 + \|u\|_{2q+2,B}^{2q+2}) ,$$

if $4q - 2 \leq 2q + 2$. The result now follows easily since
$$\int \psi |u'|^2 |u|^{2q} \leq \|u\|_{\infty,B}^{2q} \int \psi |u'|^2 .$$

The case $d = 2$ and $q = 1$ is more involved. We first have as above
$$\partial_t \left( \int \psi |\nabla u|^2 + \int \psi |u|^4 \right) \leq$$
$$C - \int \psi |\Delta u|^2 - \int \psi |u|^6 + \mathcal{O}(1) \int \frac{|\nabla \psi|^2}{\psi} |\nabla u|^2 + \mathcal{O}(1) \int \psi |\nabla u|^2 |u|^2 .$$

By integration by parts and polarization we finally arrive at
$$\partial_t \left( \int \psi |\nabla u|^2 + \int \psi |u|^4 \right) \leq$$
$$C - \int \psi |\Delta u|^2 + \mathcal{O}(1) \int \frac{|\nabla \psi|^2}{\psi} |\nabla u|^2 + \mathcal{O}(1) \int \psi |\nabla u|^2 |u|^2 .$$

We are going to apply the Gagliardo Nirenberg inequality to sub-domains of $B$ in order to deal with the presence of $\psi$ in the last term. We will also use this sub-domain decomposition to get a lower bound of the first term in term of Sobolev norms using Gårding's inequality.

We start by describing the domain decomposition. First of all we decompose $B$ into the disjoint union
$$B = B_0 \cup_{n=1}^{\infty} A_n$$
where $B_0$ is the ball of center $x_0$ with radius $99/100$, and $A_n$ is the annulus between the circles centered at $x_0$ and radius $1 - 1/(n+99)$ and $1 - 1/(n+100)$. We decompose moreover each annulus $A_n$ into a disjoint union of $[2\pi(n+100)^2]$ angular segments denoted by $A_{n,j}$ ($0 \leq j < [2\pi(n+100)^2]$), where $[\ ]$ denotes the integer part. In other words, in polar coordinates $(\rho, \theta)$, we have

$$A_{n,j} =$$
$$\left\{(\rho, \theta)\,;\, 1 - \frac{1}{n+99} > \rho \geq 1 - \frac{1}{n+100}\,,\, \frac{2\pi j}{[2\pi(n+100)^2]} \leq \theta < \frac{2\pi(j+1)}{[2\pi(n+100)^2]}\right\}\,.$$

The main point in this construction is that each $A_{n,j}$ has a diameter less than $2(n+100)^{-2}$, but it's distance to the boundary of $B$ is clearly $1/(n+100)$. Moreover, these are almost squares (and the more so if $n$ is getting large). Therefore, we can find a sequence of balls $B_{n,j}$, $0 \leq j < [2\pi(n+100)^2]$ and $n = 1, 2, \cdots$ with a radius $R_n$ equal to $2(n+100)^{-2}$, each $B_{n,j}$ covering $A_{n,j}$, and at a distance at least $1/2(n+100)$ of the boundary of $B$. From this last remark, we deduce that for some constant $c > 0$ we have

$$\sup_{x \in 2B_{n,j}} \psi(x) \leq e^c \inf_{y \in 2B_{n,j}} \psi(y)\,,$$

where $2B_{n,j}$ denotes the ball with the same center as $B_{n,j}$ but a double radius.

The reason for going from the $A_{n,j}$ to the $B_{n,j}$ is that by scaling we will be able to control the constants appearing in the Gagliardo-Nirenberg inequalities which in general depend on scale and geometry. By going to the disks we will only have to control the scale since the geometry will always be the same. The idea will be to rescale each ball $B_{n,j}$ to a ball of radius 1, then apply the corresponding Gagliardo-Nirenberg inequality, and then downscale back.

Denoting by $x_{n,j}$ the center of the ball $B_{n,j}$, we have

$$\int_{B \setminus B_0} \psi |\nabla u|^2 |u|^2 = \sum_{n,j} \int_{A_{n,j}} \psi |\nabla u|^2 |u|^2 \leq \sum_{n,j} e^c \psi(x_{n,j}) \int_{B_{n,j}} |\nabla u|^2 |u|^2\,.$$

We now bound each term separately. From the Hölder inequality we deduce

$$\int_{B_{n,j}} |\nabla u|^2 |u|^2 \leq \|\nabla u\|_{8/3, B_{n,j}}^2 \|u\|_{8, B_{n,j}}^2\,.$$

As explained before, we are now going to use scaled versions of the Gagliardo-Nirenberg inequalities. In the unit ball $B$ we have for some constant $K$ (see [F.])

$$\|v\|_{8,B} \leq K \|v\|_{1,8/3,B}^{1/3} \|v\|_{4,B}^{2/3} \quad \text{and} \quad \|\nabla v\|_{8/3,B} \leq K \|v\|_{2,2,B}^{1/2} \|v\|_{4,B}^{1/2}\,.$$

If on the other hand $B'$ is a ball of radius $R$, we have after rescaling the above ineqalities and using Hölder's inequality

$$\int_{B'} |\nabla u|^2 |u|^2 \leq$$
$$K_1 \left(\sum_{|\alpha|=2} \|D^\alpha u\|_{2,B'}^{4/3} \|u\|_{4,B'}^{8/3} + R^{-4/3} \|\nabla u\|_{2,B'}^{4/3} \|u\|_{4,B'}^{8/3} + R^{-8/3} \|u\|_{4,B'}^{8/3} \|u\|_{2,B'}^{4/3}\right)$$

where $K_1$ is independent of $R$ and $B'$. In other words, we obtain for the term we want to estimate with $R_n = 2(n+100)^{-2}$, and including the estimate for the central ball with the index $n = 0$ (and $j$ undefined)

$$\int_B \psi |\nabla u|^2 |u|^2 \leq \mathcal{O}(1) \sum_{n,j} \psi(x_{n,j}) \left( \sum_{|\alpha|=2} \|D^\alpha u\|_{2,B_{n,j}}^{4/3} \|u\|_{4,B_{n,j}}^{8/3} \right.$$

$$\left. + R_n^{-4/3} \|\nabla u\|_{2,B_{n,j}}^{4/3} \|u\|_{4,B_{n,j}}^{8/3} + R_n^{-8/3} \|u\|_{4,B_{n,j}}^{8/3} \|u\|_{2,B_{n,j}}^{4/3} \right) .$$

In order to polarize the first term of the sum against the integral of $(\Delta u)^2$, we first have to use Gårding's inequality. We will again do this using rescaling. Let $rB$ denote the ball centered at the origin and of radius $r$. Let $\xi$ be a function in $C_0^\infty(\mathbb{R}^2)$ with support in $3/2B$, such that $0 \leq \xi \leq 1$ and $\xi = 1$ on a neighborhood of $B$. It is easy to verify that

$$\int_{2B} |\Delta u|^2 \geq \int_{2B} \xi^2 |\Delta u|^2 \geq \frac{1}{2} \int_{2B} |\Delta(\xi u)|^2 - \mathcal{O}(1) \|u\|_{1,2,2B}^2 .$$

We now observe that the function $\xi u$ is in $H_0^2(2B)$, therefore we can apply Gårding's inequality (see [F.]) and we obtain

$$\int_{2B} |\Delta(\xi u)|^2 \geq \mathcal{O}(1) \int_{2B} \sum_{|\alpha|=2} |D^\alpha(\xi u)|^2 - \mathcal{O}(1) \|\xi u\|_{1,2,2B}^2 .$$

After polarization and combining the above inequalities, we finally obtain

$$\int_{2B} |\Delta u|^2 \geq C_1 \int_B \sum_{|\alpha|=2} |D^\alpha u)|^2 - \mathcal{O}(1) \|u\|_{1,2,2B}^2 ,$$

and the scaled version

$$\int_{2B'} |\Delta(u)|^2 \geq C_1 \int_{B'} \sum_{|\alpha|=2} |D^\alpha u)|^2 - C_2 \left( R^{-2} \|\nabla u\|_{2,2B'}^2 + R^{-4} \|u\|_{2,2B'}^2 \right) ,$$

where $C_1 > 0$ and $C_2 > 0$ are constants which are independent of $B'$.

We now remark that there is an integer $N$ independent of $n$ and $j$ such that each ball $2B_{n,j}$ meats at most $N$ of the domains $A_{m,l}$. Therefore, we can as before resum all the contributions, including the central one (we have to use another ball instead of $2B_0$ which is not contained in $B$, but the estimate is similar) and we finally get for some $C_3 > 0$

$$\int_B \psi |\Delta u|^2 \geq C_3 \sum_{n,j} \psi(x_{n,j}) \int_{B_{n,j}} \sum_{|\alpha|=2} |D^\alpha u|^2$$

$$- \mathcal{O}(1) \sum_{n,j} \psi(x_{n,j}) \left( R_n^{-2} \|\nabla u\|_{2,2B_{n,j}}^2 + R_n^{-4} \|u\|_{2,2B_{n,j}}^2 \right) .$$

Therefore, we finally obtain for some numbers $a > 0$ and $b > 0$

$$-\int \psi |\Delta u|^2 + \mathcal{O}(1) \int \psi |\nabla u|^2 |u|^2$$

$$\leq \sum_{n,j} \psi(x_{n,j}) \sum_{|\alpha|=2} (-a\|D^\alpha u\|_{2,B_{n,j}} + b\|D^\alpha u\|_{2,B_{n,j}}^{4/3} \|u\|_{4,B_{n,j}}^{8/3})$$

$$+ \sum_{n,j} \psi(x_{n,j}) \left( R_n^{-2} \|\nabla u\|_{2,2B_{n,j}}^2 + R_n^{-4} \|u\|_{2,2B_{n,j}}^2 \right.$$

$$+ R_n^{-4/3} \|\nabla u\|_{2,B_{n,j}}^{4/3} \|u\|_{4,B_{n,j}}^{8/3} + R_n^{-8/3} \|u\|_{4,B_{n,j}}^{8/3} \|u\|_{2,B_{n,j}}^{4/3} \right) .$$

We now polarize each term for a fixed pair $(n,j)$ and we get

$$-\int \psi |\Delta u|^2 + \mathcal{O}(1) \int \psi |\nabla u|^2 |u|^2$$

$$\leq \mathcal{O}(1) \sum_{n,j} \psi(x_{n,j}) \|u\|_{4,B_{n,j}}^8 + \sum_{n,j} \psi(x_{n,j}) \left( R_n^{-2} \|\nabla u\|_{2,2B_{n,j}}^2 + R_n^{-4} \|u\|_{2,2B_{n,j}}^2 \right) .$$

We now observe that because of our choice of $\psi$, there is a uniform constant $C_3$ such that for any $n$ and $j$ we have

$$\psi(x_{n,j}) R_n^{-4} \leq C_3 .$$

Therefore we finally obtain using the obvious estimate

$$\|u\|_{4,B_{n,j}}^8 \leq \|u\|_{4,B}^4 \|u\|_{4,B_{n,j}}^4$$

and reinserting $\psi$ into the integrals

$$-\int \psi |\Delta u|^2 + \mathcal{O}(1) \int \psi |\nabla u|^2 |u|^2 \leq \mathcal{O}(1) \int \psi |u|^4 \int_B |u|^4 + \mathcal{O}(1) \int_B (|\nabla u|^2 + |u|^2) .$$

The proof is completed by inserting the results from Theorem 1.

We observe that by letting $\alpha$ and $\beta$ tend to infinity with a finite fixed negative ratio and rescaling the equation we obtain formally the non linear Schrödinger equation which is known to have singularities in dimension larger than 1 (see [K.] for the existence theorem). It is of course an interesting and open question to know if blowing up solutions exist for $CGL_3$ in any dimension (larger than 2).

We now state two immediate but useful corollaries.

**Corollary 5.** *Under the hypothesis of Theorem 3, there is a positive constant $C$ such that for any $0 < \tau < T$ we have*

$$\int \varphi |\nabla u_T|^2 + \rho \int \varphi |u_T|^{2q+2} \leq C(\tau + 1 + y(T - \tau)(1 + 1/\tau)) .$$

**Proof.** This is an immediate consequence of Theorem 3. Indeed, let $z(t)$ be the function defined by

$$z(t) = \int \varphi |\nabla u_t|^2 + \rho \int \varphi |u_t|^{2q+2} + \theta \int \left(\varphi + \frac{|\nabla \varphi|^2}{\varphi}\right) |u_t|^2 ,$$

then from Theorem 3 we have $\partial_t z \leq C$, therefore if $0 < T - \tau < s < T$ we have

$$z(T) \leq z(s) + C(T - s) .$$

The result follows by averaging this inequality over s in the interval $[T - \tau, T]$ and using Corollary 2.

One could of course optimize the previous estimate with respect to $\tau$, we will prefer to keep it in this form in view of the application given in the next section.

**Corollary 6.** *Under the hypothesis of Theorem 4, there are positive constants $C$ and $D$ such that the quantity*

$$z(t) = \int \psi |\nabla u|^2 + \int \psi |u|^{2q+2} + \rho \int \zeta |u|^2$$

*satisfies for any $0 < \tau < T$ the estimate*

$$z(T) \leq C \left(\frac{y(T - \tau)}{\tau} + 1 + \tau\right) e^{D(\tau + y(T-\tau))} .$$

**Proof.** This follows from Theorem 4 by a well known estimation, we sketch the proof here for the convenience of the reader. First of all let $z_1(t)$ be the function

$$z_1(t) = \int \zeta |u|^4 ,$$

then if we define the function $v(t)$ (for $T - \tau \leq t \leq T$) by

$$v(t) = z(t) e^{-c \int_{T-\tau}^{t} z_1(s) ds}$$

we have for $u$ the differential inequality (since $z_1 \geq 0$)

$$\partial_t v \leq C .$$

It follows at once using Corollary 2 that for $T - \tau \leq s \leq t \leq T$ we have

$$z(t) \leq (z(s) + C(t - s)) e^{D(\tau + y(T-\tau))} .$$

The result follows by averaging over $s$ on the interval $[T - \tau, T]$ and using Corollary 2.

We make the important remark that the two bounds obtained in corollaries 5 and 6 are non increasing in $T$ for a fixed $\tau$ since they depend explicitly on the function $y$. Moreover, if $T$ can be taken large enough (in particular larger than one), these estimates become independent of the initial condition.

# III $L^\infty$ ESTIMATES.

In this section we will use the integral equations and the local energy estimates in order to derive $L^\infty$ estimates uniform in time. We first introduce some notations and recall some well known results.

Let $G_s$ denote the kernel of the operator $\exp(s(1+i\alpha)\Delta)$ ($s > 0$). From translation invariance, this kernel is a function of the difference of it's arguments, denoted below by $g_t$, namely
$$G_t(x, y) = g_t(x - y) .$$

Moreover, this function is even. Using Fourier transform it is easy to verify that there is a positive constant $c$ which depends only on the space dimension $d$ and the number $\alpha$ such that
$$|G_s(x,y)| = |g_t(x-y)| \leq c^{-1} s^{-d/2} e^{-c|x-y|^2/s} .$$

Using Fourier transform, one can easily get an explicit expression of $c$ in terms of $\alpha$.

We deduce that for any numbers $T > \tau > 0$

$$u_T = g_\tau * u_{T-\tau} + \int_{T-\tau}^{T} ds \, g_{T-s} * NL(u, \overline{u}) . \tag{3.1}$$

where $*$ denotes convolution (in space), and $u_s$ is the function $u(\cdot, s)$. Note that this formula makes sense if we know that the solution exists and is in $L^\infty$ uniformly on the interval $[T - \tau, T]$.

Although we could derive the $L^\infty$ bound using this formula, we will use a slightly different expression which is localized in space. This will simplify slightly the analysis but also provides immediatly an estimate valid for large but finite domains. Let $\zeta$ be a $C^\infty$ function with compact support, which is non negative and equal to 1 on the ball $Q_1$ of radius one centered at the origin. As usual, we will in fact use translates of this function, and for convenience we will denote by $\zeta_x$ the function $\zeta(\cdot - x)$. We will denote by $Q$ the support of $\zeta$.

**Lemma 7.** *For any number $T > 0$ such that the solution of $CGL_q$ is in $L^\infty$ on the interval $[0, T]$, and for any number $0 < \tau < T$ we have*

$$u_T(x) = g_\tau * (\zeta_x u_{T-\tau}) +$$

$$\int_{T-\tau}^{T} ds \, g_s * (\zeta_x NL(u_{T-s}, \overline{u}_{T-s}) - (1+i\alpha) u_{T-s} \Delta \zeta_x - (1+i\alpha) \nabla \zeta_x \bullet \nabla u_{T-s}) . \tag{3.2}$$

The proof is a standard argument writing the function $g_t * (\zeta_x u_{T-t})$ as the integral of its derivative with respect to $t$ and using the equation $CGL_q$.

The strategy of the estimate will be as follows. First we have the local time existence Theorem which we recall now for the convenience of the reader.

**Theorem 8.** *Let $u_0$ be a function in $L^\infty$. Then there is a number $1 > T_0 > 0$ such that on the interval $[0, T_0]$, $CGL_q$ has a unique solution with initial condition $u_0$. This solution satisfies for a constant $C > 1$ independent of $u_0$*

$$\sup_{t \in [0, T_0]} \|u_t\|_\infty < C \|u_0\|_\infty \,,$$

*and*

$$\sup_{t \in [T_0/2, T_0]} (\|\Delta u_t\|_\infty + \|\nabla u_t\|_\infty) < C T_0^{1/2} \|u_0\|_\infty \,.$$

*Moreover, we have*

$$T_0 \geq C^{-1} \|u_0\|_\infty^{-2q} \,.$$

We refer to [P.] and [H.] for a proof. The number $T_0$ may be larger than 1 (for small initial data for example) but for the applications below it is usefull to introduce a uniform upper bound, and we will always assume $T_0 < 1$.

Now for a given time $T$ such that

$$\sup_{t \in [0, T]} \|u_t\|_\infty < \infty \,,$$

we will chose a number $\tau < \inf(T/2, 1)$. This number will depend on $T$ in particular but will be bounded below away from zero and non decreasing in $T$. It will moreover converge to 1 for large $T$. In fact we can take $\tau = 1$ as soon as $T > 2$. We will proove a bound for $\|u_T\|_\infty$ given by

$$\|u_T\|_\infty \leq \frac{1}{2} \|u_{T-\tau}\|_\infty + F(T, \tau) \tag{3.3}$$

where the function $F$ depends on $T$ only through the function $y$ of Corollary 2. It will follow from this explicit dependence that for a fixed $\tau$, this function will be non increasing in $T$ with a limit reached for a finite $T$. The first result in this direction is of course the local time existence theorem above, it is first used to produce the initial $T = T_0$, and then to increase this number by a finite amount at each step. More precisely, we will start with an initial existence time $T_0$, and take for example $\tau = T_0/2$ as long as $T < 2$. Using (3.3), we derive an estimate on the $L^\infty$ norm which will never deteriorate with time, namely

$$\|u_T\|_\infty \leq \|u_0\|_\infty + 2F(T, T_0/2) \leq \|u_0\|_\infty + 2F(T_0, T_0/2) \,.$$

We can now use Theorem 8 to increase the interval of existence in $L^\infty$ from $[0, T]$ to $[0, T + \delta]$ where $\delta$ can be taken independent of $T$ using the estimate of Theorem 8. We then repeat the estimate coming from (3.3) on this new larger interval, and proceeds like this until $T$ reaches 2. At this stage we take for ever the value $\tau = 1$ and continue the extension of the exitence interval. As before, the $L^\infty$ bound will never deteriorate and therefore we can continue for ever. Moreover, for fixed $\tau$ $(= 1)$ the $L^\infty$ bound depends on $T$ only through the function $y$ of Corollary 2, therefore there is a finite time $T$ after which our bound stabilizes to a value independent of the initial condition. We now state and prove the $L^\infty$ bound.

**Theorem 9.** *For any $CGL_q$ in dimension 1, 2 or 3 satisfying the hypothesis of Thoerems 3 or 4, the solution $u$ of the equation is uniformly bounded in time and space. There is moreover a ball in $L^\infty$ which is globally absorbing.*

**Proof.** In order to simplify the notations, we will assume that $P = 0$. The general case is treated by obvious minor modifications of the folowing arguments. The case of dimension 1 is particularly simple, since the $L^\infty$ estimates follows at once from Corollaries 2, 4 and 5 using a Sobolev inequality. We now only consider the case of dimensions 2 and 3.

From the translation invariance of the problem, it is enough to estimate the function $u$ at the origin. As explained before we will assume that the solution exists in $L^\infty$ on the time interval $[0, T]$.

We will denote by $Q$ the support of the function $\zeta$, and by $Q_l$ the unit ball centered at the origin with radius $l$.

For a fixed number $\tau > 0$ smaller than one, we choose a number $l < 1/3$ small enough such that

$$\int_{Q_l} |g_\tau(y)| \, d^d y \leq 1/2 \ .$$

We now observe that there is a constant $A > 0$ such that if $x$ and $y$ are at a distance larger than $l/3$, we have for $0 < s \leq \tau$

$$|g_s(x - y)| \leq A \ .$$

Note that $A$ depends on $\tau$. Therefore, if $0 < \tau < T$ and $\tau \leq 1$ we have

$$\left| \int_{T-\tau}^T ds \int g_{T-s} \zeta u_s |u_s|^{2q} \right| \leq$$

$$\int_{T-\tau}^T ds \int_{Q_l} g_{T-s} |u_s|^{2q+1} + \mathcal{O}(1) A \int_{T-\tau}^T ds \int_{Q \setminus Q_l} |u_s|^{2q+1} \ .$$

Since $2q + 1 < 2q + 2$, the second sum is uniformly bounded from Corollary 2, and the bound can be taken independent of $u$ if $T$ is large enough. For the central term we use the Hölder inequality to get

$$\left| \int_{Q_l} g_{T-s} u_s |u_s|^{2q} \right| \leq \mathcal{O}(1)(T-s)^{-d(2q+1)/4(q+1)} \|u_s\|_{2q+2, Q_l}^{2q+1} \ .$$

We now use the uniform bounds of Theorems 3 and 4. If $d = 2$ the above quantity is integrable in $s$ and this leads to a uniform bound. Finally we can estimate the last terms in equation (3.2) by the same method. This is even simpler since from the properties of $\zeta$ the integration is only over the set $Q \setminus Q_1$.

For $d = 3$ and $q = 1$, we first observe that Theorem 4 provides in fact a better bound. Indeed, using the Gagliardo-Nirenberg inequality in the domain $Q_l$ we have

$$\|u_s\|_{6, Q_l} \leq \mathcal{O}(1) \|u_s\|_{1, 2, Q_l} \ .$$

Therefore

$$\left| \int_{Q_l} g_{T-s} u_s |u_s|^2 \right| \leq \mathcal{O}(1)(T-s)^{-3/4} \|u_s\|_{6,Q_l}^3 \leq \mathcal{O}(1)(T-s)^{-3/4} \|u_s\|_{1,2,Q_l}^3 \ .$$

From thereon, the argument proceeds as before.

At this point, using Corollaries 5 and 6, we can construct a function $F_1(T,\tau)$ which depends on $T$ only through the function $y$ of Corollary 2, and such that for $T > \tau > 0$ we have

$$\int_{T-\tau}^T ds \int g_{T-s} \left( |u_s| |\Delta \zeta_x| + |\nabla \zeta_x| |\nabla u_s| \right) + \left| \int_{T-\tau}^T ds \int g_{T-s} u_s |u_s|^{2q} \right| \leq F_1(T,\tau) \ .$$

Moreover, it is easy to verify that for a fixed $\tau$ this function is not increasing, and it reaches a limit after some finite $T$, this limit being independent of the initial condition.

Finally, we have to deal with the first term in equation (3.2). We decompose the integral into a sum of two integrals with domain $Q_l$ and $Q \backslash Q_l$ respectively. The integral over $Q \backslash Q_l$ is controled as above. This provides a second function $F_2$ with the same properties as $F_1$, and we set $F = F_1 + F_2$. It therefore remains to investigate the central term which is given by

$$\int_{Q_l} g_\tau u_{T-\tau} \ .$$

From our choice of $l$ it follows that

$$\left| \int_{Q_l} g_\tau u_{T-\tau} \right| \leq \frac{1}{2} \|u_{T-\tau}\|_\infty \ ,$$

which implies, using translation invariance and the previous estimates

$$\|u_T\|_\infty \leq \frac{1}{2} \|u_{T-\tau}\|_\infty + F(T,\tau) \ .$$

As explained before, this immediately leads by iteration to the proof of the theorem.

We now consider briefly the problem $\text{CGL}_q$ in bounded domain. All the bounds which have been established before are of local nature, therefore they are also valid in a bounded domain. The only restriction is that one should be at some distance from the boundary. This distance does not depend on the size of the domain nor on the boundary condition, it depends only on the coefficients of the equation. A simple consequence of this result is the following. Assume that we are in the conditions of Theorem 9 in a finite domain $\Omega$ large enough and fix the boundary condition by imposing that at the boundary the function $u$ is equal to some constant $a$ (Dirichlet type condition). Assume now that $|a|$ is much larger than the radius of the absorbing $L^\infty$ ball of Theorem 9. Then starting with a bounded initial condition, a boundary layer develops where the solution interpolates between the boundary value $a$ and values inside the attracting ball. The width of this boundary layer depends only on the coefficients of the equation and not on $a$ or on the size of the domain. Note also that this stabilization of the bound outside the boundary layer is reached after a finite time.

**References.**